\documentstyle[twoside,epsfig]{article}

\catcode`\@=11
\long\def\@makefntext#1{
\protect\noindent \hbox to 3.2pt {\hskip-.9pt  
$^{{\eightrm\@thefnmark}}$\hfil}#1\hfill}		%CAN BE USED 

\def\thefootnote{\fnsymbol{footnote}}
\def\@makefnmark{\hbox to 0pt{$^{\@thefnmark}$\hss}}	%ORIGINAL 
	
\def\ps@myheadings{\let\@mkboth\@gobbletwo
\def\@oddhead{\hbox{}
\rightmark\hfil\eightrm\thepage}   
\def\@oddfoot{}\def\@evenhead{\eightrm\thepage\hfil
\leftmark\hbox{}}\def\@evenfoot{}
\def\sectionmark##1{}\def\subsectionmark##1{}}

\oddsidemargin=\evensidemargin
\renewcommand{\thefootnote}{\fnsymbol{footnote}}

\newcounter{sectionc}\newcounter{subsectionc}\newcounter{subsubsectionc}
\renewcommand{\section}[1] {\vspace{12pt}\addtocounter{sectionc}{1} 
\setcounter{subsectionc}{0}\setcounter{subsubsectionc}{0}\noindent 
	{\tenbf\thesectionc. #1}\par\vspace{5pt}}
\renewcommand{\subsection}[1] {\vspace{12pt}\addtocounter{subsectionc}{1} 
	\setcounter{subsubsectionc}{0}\noindent 
	{\bf\thesectionc.\thesubsectionc. {\kern1pt \bfit #1}}\par\vspace{5pt}}
\renewcommand{\subsubsection}[1] {\vspace{12pt}\addtocounter{subsubsectionc}{1}
	\noindent{\tenrm\thesectionc.\thesubsectionc.\thesubsubsectionc.
	{\kern1pt \tenit #1}}\par\vspace{5pt}}
\newcommand{\nonumsection}[1] {\vspace{12pt}\noindent{\tenbf #1}
	\par\vspace{5pt}}

\newcounter{appendixc}
\newcounter{subappendixc}[appendixc]
\newcounter{subsubappendixc}[subappendixc]
\renewcommand{\thesubappendixc}{\Alph{appendixc}.\arabic{subappendixc}}
\renewcommand{\thesubsubappendixc}
	{\Alph{appendixc}.\arabic{subappendixc}.\arabic{subsubappendixc}}

\renewcommand{\appendix}[1] {\vspace{12pt}
        \refstepcounter{appendixc}
        \setcounter{figure}{0}
        \setcounter{table}{0}
        \setcounter{lemma}{0}
        \setcounter{theorem}{0}
        \setcounter{corollary}{0}
        \setcounter{definition}{0}
        \setcounter{equation}{0}
        \renewcommand{\thefigure}{\Alph{appendixc}.\arabic{figure}}
        \renewcommand{\thetable}{\Alph{appendixc}.\arabic{table}}
        \renewcommand{\theappendixc}{\Alph{appendixc}}
        \renewcommand{\thelemma}{\Alph{appendixc}.\arabic{lemma}}
        \renewcommand{\thetheorem}{\Alph{appendixc}.\arabic{theorem}}
        \renewcommand{\thedefinition}{\Alph{appendixc}.\arabic{definition}}
        \renewcommand{\thecorollary}{\Alph{appendixc}.\arabic{corollary}}
        \renewcommand{\theequation}{\Alph{appendixc}.\arabic{equation}}
        \noindent{\tenbf Appendix \theappendixc #1}\par\vspace{5pt}}
\newcommand{\subappendix}[1] {\vspace{12pt}
        \refstepcounter{subappendixc}
        \noindent{\bf Appendix \thesubappendixc. {\kern1pt \bfit #1}}
	\par\vspace{5pt}}
\newcommand{\subsubappendix}[1] {\vspace{12pt}
        \refstepcounter{subsubappendixc}
        \noindent{\rm Appendix \thesubsubappendixc. {\kern1pt \tenit #1}}
	\par\vspace{5pt}}

\newcommand{\textlineskip}{\baselineskip=13pt}
\newcommand{\smalllineskip}{\baselineskip=10pt}

\def\eightcirc{
\begin{picture}(0,0)
\put(4.4,1.8){\circle{6.5}}
\end{picture}}
\def\eightcopyright{\eightcirc\kern2.7pt\hbox{\eightrm c}} 

\newcommand{\copyrightheading}[1]
	{\vspace*{-2.5cm}\smalllineskip{\flushleft
	{\footnotesize International Journal of Modern Physics A, #1}\\
	{\footnotesize $\eightcopyright$\, World Scientific Publishing
	 Company}\\
	 }}

\def\abstracts#1#2#3{{
	\centering{\begin{minipage}{4.5in}\baselineskip=10pt\footnotesize
	\parindent=0pt #1\par 
	\parindent=15pt #2\par
	\parindent=15pt #3
	\end{minipage}}\par}} 

\renewenvironment{thebibliography}[1]
	{\frenchspacing
	 \ninerm\baselineskip=11pt
	 \begin{list}{\arabic{enumi}.}
	{\usecounter{enumi}\setlength{\parsep}{0pt}
	 \setlength{\leftmargin 12.7pt}{\rightmargin 0pt} %FOR 1--9 ITEMS
	 \setlength{\itemsep}{0pt} \settowidth
	{\labelwidth}{#1.}\sloppy}}{\end{list}}

\newcounter{itemlistc}
\newcounter{romanlistc}
\newcounter{alphlistc}
\newcounter{arabiclistc}

\newcommand{\fcaption}[1]{
        \refstepcounter{figure}
        \setbox\@tempboxa = \hbox{\footnotesize Fig.~\thefigure. #1}
        \ifdim \wd\@tempboxa > 5in
           {\begin{center}
        \parbox{5in}{\footnotesize\smalllineskip Fig.~\thefigure. #1}
            \end{center}}
        \else
             {\begin{center}
             {\footnotesize Fig.~\thefigure. #1}
              \end{center}}
        \fi}

\newcommand{\tcaption}[1]{
        \refstepcounter{table}
        \setbox\@tempboxa = \hbox{\footnotesize Table~\thetable. #1}
        \ifdim \wd\@tempboxa > 5in
           {\begin{center}
        \parbox{5in}{\footnotesize\smalllineskip Table~\thetable. #1}
            \end{center}}
        \else
             {\begin{center}
             {\footnotesize Table~\thetable. #1}
              \end{center}}
        \fi}

\def\@citex[#1]#2{\if@filesw\immediate\write\@auxout
	{\string\citation{#2}}\fi
\def\@citea{}\@cite{\@for\@citeb:=#2\do
	{\@citea\def\@citea{,}\@ifundefined
	{b@\@citeb}{{\bf ?}\@warning
	{Citation `\@citeb' on page \thepage \space undefined}}
	{\csname b@\@citeb\endcsname}}}{#1}}

\newif\if@cghi
\def\cite{\@cghitrue\@ifnextchar [{\@tempswatrue
	\@citex}{\@tempswafalse\@citex[]}}
\def\citelow{\@cghifalse\@ifnextchar [{\@tempswatrue
	\@citex}{\@tempswafalse\@citex[]}}
\def\@cite#1#2{{$\null^{#1}$\if@tempswa\typeout
	{IJCGA warning: optional citation argument 
	ignored: `#2'} \fi}}

\def\pmb#1{\setbox0=\hbox{#1}
	\kern-.025em\copy0\kern-\wd0
	\kern.05em\copy0\kern-\wd0
	\kern-.025em\raise.0433em\box0}

\def\fnm#1{$^{\mbox{\scriptsize #1}}$}
\def\fnt#1#2{\footnotetext{\kern-.3em
	{$^{\mbox{\scriptsize #1}}$}{#2}}}

\def\fpage#1{\begingroup
\voffset=.3in
\thispagestyle{empty}\begin{table}[b]\centerline{\footnotesize #1}
	\end{table}\endgroup}

\def\runninghead#1#2{\pagestyle{myheadings}
\markboth{{\protect\footnotesize\it{\quad #1}}\hfill}
{\hfill{\protect\footnotesize\it{#2\quad}}}}
\font\tenrm=cmr10
\font\tenit=cmti10 
\font\tenbf=cmbx10
\font\bfit=cmbxti10 at 10pt
\font\ninerm=cmr9

\font\eightrm=cmr8

\def\qed{\hbox{${\vcenter{\vbox{			%HOLLOW SQUARE
   \hrule height 0.4pt\hbox{\vrule width 0.4pt height 6pt
   \kern5pt\vrule width 0.4pt}\hrule height 0.4pt}}}$}}

\renewcommand{\thefootnote}{\fnsymbol{footnote}}	%USE SYMBOLIC FOOTNOTE
\def\tdzresult{414.8\pm3.8\pm3.4}
\def\tdpresult{1040^{+23}_{-22}\pm18}
\def\tdsresult{479^{+17}_{-16}{}^{+6}_{-8}}
\def\rdpresult{2.51\pm 0.06\pm0.04}
\def\rdsresult{1.15\pm 0.04^{+0.01}_{-0.02}}
\def\yresult{1.0^{+3.8}_{-3.5}{}^{+1.1}_{-2.1}}
\def\yllimit{-7.0}
\def\yulimit{8.7}
\def\intl{2.75}

\newcommand{\prl}[3]{Phys. Rev. Lett. {\bf #1} (#3) #2}

\newcommand{\plb}[3]{Phys. Lett. B {\bf #1} (#3) #2}

\newcommand{\zpc}[3]{Z. Phys. C {\bf #1} (#3) #2}

\def\tdz{\tau(\dz)}
\def\tdp{\tau(\dplus)}
\def\tds{\tau(\ds)}
\def\rdp{\tdp/\tdz}
\def\rds{\tds/\tdz}
\def\Tdzresult{$\tdz=(\tdzresult)$ fs}
\def\Tdpresult{$\tdp=(\tdpresult)$ fs}
\def\Tdsresult{$\tds=(\tdsresult)$ fs}
\def\Rdpresult{$\rdp=\rdpresult$}
\def\Rdsresult{$\rds=\rdsresult$}
\def\Yresult{$\ycp=(\yresult)$ \%}
\def\Ylimit{$\yllimit\%<\ycp<\yulimit$\%}
\def\etal{{\it et al.}}
\def\ycp{y_{CP}}
\def\am{A_{mix}}

\def\Fb{fb$^{-1}$}

\def\dzdzbar{D^0\overline{D}^0}

\def\DzDzbar{$\dzdzbar$}

\def\pip{\pi^+}
\def\pim{\pi^-}
\def\piz{\pi^0}
\def\km{K^-}
\def\kp{K^+}
\def\kstarzb{\overline{K}^{*0}}

\def\dstarp{D^{*+}}

\def\dz{D^0}
\def\dplus{D^+}
\def\ds{D_s^+}
\def\Dz{$\dz$}
\def\Dp{$\dplus$}
\def\Ds{$\ds$}

\def\kpkm{\kp\km}
\def\kmkp{\km\kp}
\def\kpi{\km\pip}

\def\kpipi{\km\pip\pip}

\def\dzkpi{\dz\to\kpi}
\def\dzkk{\dz\to\kmkp}
\def\dpphipi{\dplus\to\phi\pip}
\def\dpkpipi{\dplus\to\kpipi}
\def\dsphipi{\ds\to\phi\pip}
\def\dskstk{\ds\to\kstarzb\kp}

\def\vec2#1{\mbox{\boldmath $#1$}}
\begin{document}

\runninghead{Measurement of Charmed Meson Lifetimes with Belle}
{Measurement of Charged Meson Lifetimes with Belle}

\normalsize\textlineskip
\thispagestyle{empty}
\setcounter{page}{1}

\copyrightheading{}			%{Vol. 0, No. 0 (1993) 000--000}

\vspace*{0.88truein}

\fpage{1}
\centerline{\bf MEASUREMENT OF CHARMED MESON LIFETIMES WITH BELLE}
%\vspace*{0.035truein}
%\centerline{\bf WITH BELLE}
\vspace*{0.37truein}
\centerline{\footnotesize JUN-ICHI TANAKA\footnote{
jtanaka@hep.phys.s.u-tokyo.ac.jp
}}
\vspace*{0.015truein}
\centerline{\footnotesize\it Department of Physics,
University of Tokyo,
7-3-1 Hongo}
\baselineskip=10pt
\centerline{\footnotesize\it Bunkyo-ku, Tokyo 113-003, Japan}
%\vspace*{0.225truein}
%\publisher{(received date)}{(revised date)}

\vspace*{0.21truein}
\abstracts{The lifetimes of charmed mesons have been measured
using \intl~\Fb\ of data collected with the Belle detector at KEKB. 
Each candidate is fully reconstructed to identify the flavor of the charmed meson. 
The lifetimes are measured to be \Tdzresult, \Tdpresult\ and \Tdsresult,
where the first error is statistical and the second error is systematic. 
The ratios of the lifetimes of \Dp\ and \Ds\ with respect to \Dz\ are
measured to be \Rdpresult\ and \Rdsresult. 
The mixing parameter $\ycp$ is also measured through the lifetime
difference of \Dz\ mesons decaying into CP-mixed states and CP eigenstates.
We find \Yresult, corresponding to a 95\% confidence interval \Ylimit.
All results are preliminary.
}{}{}

%\textlineskip			%) USE THIS MEASUREMENT WHEN THERE IS
%\vspace*{12pt}			%) NO SECTION HEADING

\vspace*{1pt}\textlineskip	%) USE THIS MEASUREMENT WHEN THERE IS

%%\section{Introduction}	        %) A SECTION HEADING
%%\vspace*{-0.5pt}
%%\noindent
\vspace*{0.21truein}
Measurements of individual charmed meson lifetimes provide
useful information for the theoretical understanding of
the heavy flavor decay mechanisms\cite{life-th}$^,$\cite{life-ratio-th}.
Moreover, the \DzDzbar\ mixing parameters,
$y\equiv(\Gamma_H-\Gamma_L)/{(\Gamma_H+\Gamma_L)}$ and
$x\equiv 2(M_H-M_L)/(\Gamma_H+\Gamma_L)$, can be explored by 
measuring the lifetime difference of the \Dz\ meson decaying
into a CP-mixed state $\dzkpi$ and a CP-eigenstate $\dzkk$.
The parameter $\ycp$, defined by
$\ycp\equiv\frac{\Gamma(\mathrm{CP\ even})-\Gamma(\mathrm{CP\ odd})}
{\Gamma(\mathrm{CP\ even})+\Gamma(\mathrm{CP\ odd})}=\frac{\tau(\dzkpi)}{\tau(\dzkk)}-1,$
is related to $y$ and $x$ by the expression
$\ycp=y\cos\phi-\frac{\am}{2} x\sin\phi,$
where $\phi$ is a CP violating weak phase due to the
interference of decays with and without mixing, and $\am$ is
related to CP violation in mixing.
E791\cite{E791a}$^,$\cite{E791b} and FOCUS\cite{FOCUS} have measured
$\ycp=(0.8\pm2.9\pm1.0)$\% and $\ycp=(3.42\pm1.39\pm0.74)$\% respectively.
It is interesting that the FOCUS result is non-zero by more
than two standard deviations.
On the other hand, CLEO\cite{CLEO} gives results for \DzDzbar\ mixing
through $\dz\to\kp\pim$, $y'\cos\phi=(-2.5^{+1.4}_{-1.6})$\%,
$x'=(0.0\pm1.5\pm0.2)$\% and $\am=0.23^{+0.63}_{-0.80}$, where
$y'=y\cos\delta - x\sin\delta$ and $x'=x\cos\delta + y\sin\delta$;
the parameter $\delta$ is the strong phase between the doubly
Cabibbo suppressed decay $D^0 \to K^+ \pi^-$ and the Cabibbo
allowed decay $\overline{D}^0 \to K^+ \pi^-$ ($\delta=0$ in the $SU(3)$ limit). 
The FOCUS and CLEO results could be consistent if there is
a large $SU(3)$ breaking effect in $\dz\to K^\pm\pi^\mp$ decays\cite{mix-th}.
%\pagebreak

\textheight=7.8truein
\setcounter{footnote}{0}
\renewcommand{\thefootnote}{\alph{footnote}}

%%\section{Reconstruction}
%%\noindent
In the lifetime measurements\cite{Belle}, 
\Dz, \Dp\ and \Ds\ mesons are fully reconstructed
via the decay chains\fnm{a}\fnt{a}{Charge-conjugate modes are implied throughout this paper.},
$\dz\to\kpi$,
$\dz\to\kmkp$,
$\dplus\to\kpipi\ ({\mathrm with}\ \dstarp\to\dplus\piz\ {\mathrm requirement})$,
$\dplus\to\phi\pip$, $\phi\to\kpkm$,
$\ds\to\phi\pip$, and
$\ds\to\kstarzb\kp$, $\kstarzb\to\kpi$.

The decay vertex($\vec2{x}_{dec}$) of the charmed meson is determined and then
the production vertex($\vec2{x}_{pro}$) is obtained by extrapolating the $D$
flight path to the interaction region of $e^+e^-$.
The projected decay length($L$) and the proper-time($t$) are obtained from
$L = (\vec2{x}_{pro}-\vec2{x}_{dec})\cdot\vec2{p}_D/|\vec2{p}_D|$ and
$t = Lm_D/c|\vec2{p}_D|$ respectively, where $\vec2{p}_D$ and $m_D$
are momentum and mass of the charmed meson.

\begin{figure}[htbp]
\vspace*{13pt}
%%\centerline{\vbox{\hrule width 5cm height0.001pt}}
%\vspace*{1.4truein}		%ORIGINAL SIZE=1.6TRUEIN x 100% - 0.2TRUEIN
\includegraphics[scale=0.3,clip]{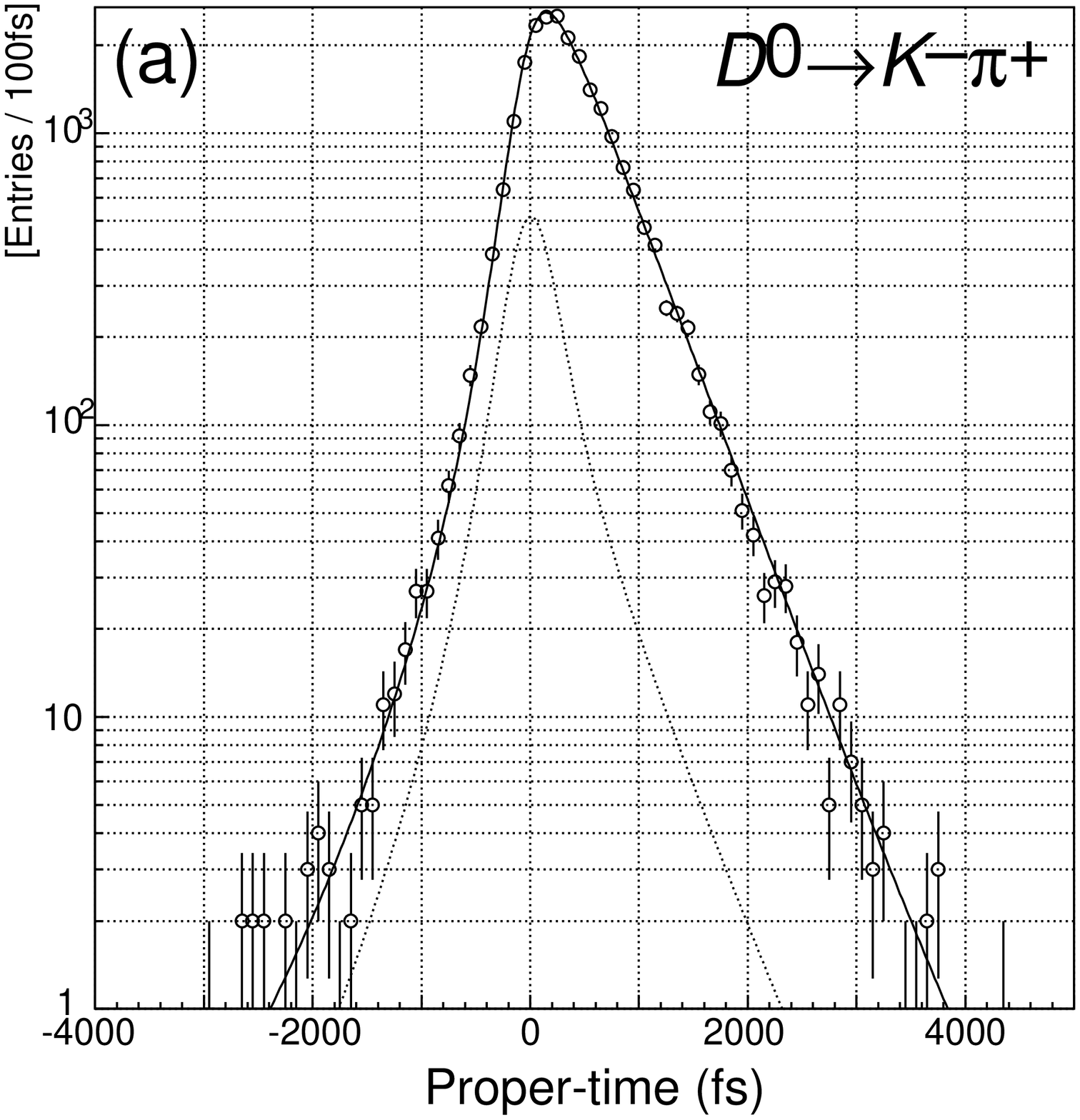}
\includegraphics[scale=0.32,clip]{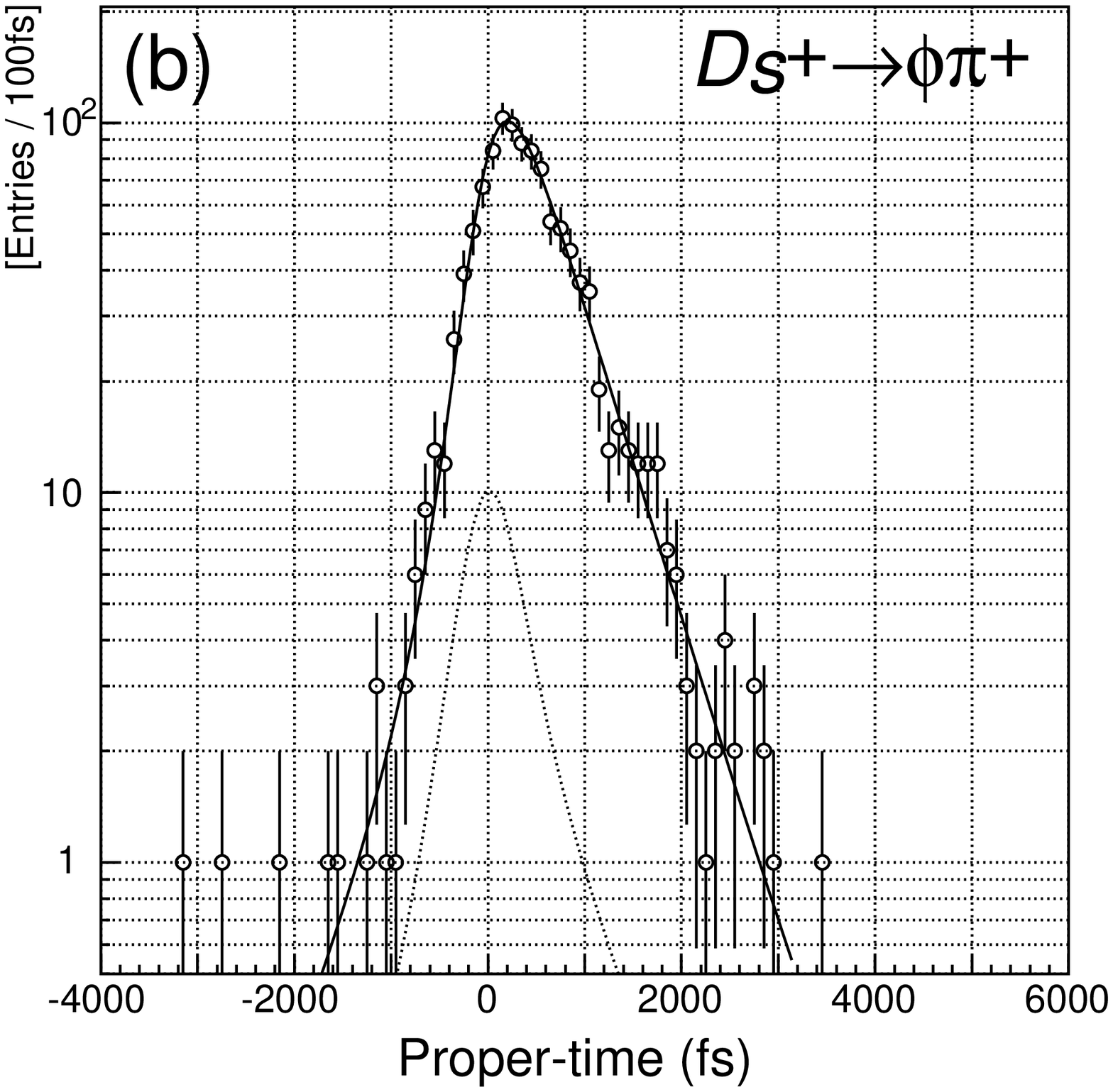}
%%\centerline{\vbox{\hrule width 5cm height0.001pt}}
\vspace*{13pt}
\fcaption{The proper-time distributions and fit results
for $\dz\to\kpi$ and $\ds\to\phi\pip$.
The dotted curve represents the background.}
\label{Fit-Results}
\end{figure}

%%\section{Lifetime Fit}
%%\noindent
An unbinned maximum likelihood fit is performed to extract the lifetimes.
The probability density function($P$) for each event consists of a signal term and
the two background terms, representing components of the background
with non-zero lifetime and zero lifetime respectively.
The likelihood function($L$) is then given by
\begin{eqnarray*}
L = \prod_i P(t^i, \sigma_t^i, f_{SIG}^i) &=&
\prod_i [f_{SIG}^i\int^\infty_0 dt^\prime\frac{1}{\tau_{SIG}}e^{\frac{-t^\prime}{\tau_{SIG}}}
R_{SIG}(t^i-t^\prime,\sigma_t^i) \\
&+&
(1-f_{SIG}^i)\int^\infty_0 dt^\prime
\{f_{\tau_{BG}}\frac{1}{\tau_{BG}}e^{\frac{-t^\prime}{\tau_{BG}}} \\
&+&(1-f_{\tau_{BG}})\delta(t^\prime)\}
R_{BG}(t^i-t^\prime,\sigma_t^i)],
\end{eqnarray*}
where $f_{SIG}^i$ and $f_{\tau_{BG}}$ are fractions for the signal and
the background with lifetime,
$\tau_{SIG}$ and $\tau_{BG}$ are the signal and background lifetimes,
$R_{SIG}$ and $R_{BG}$ are the resolution functions for the signal and
the background,
and $t^i$, $\sigma_t^i$ are the measured proper-time, and its error, for each event.
The fraction $f_{SIG}^i$ is obtained based on the charmed meson mass for each event.
The resolution functions $R_{SIG}$ and $R_{BG}$ are represented using
\begin{eqnarray*}
R(t,\sigma_t) = (1-f_{tail})\frac{1}{\sqrt{2\pi}S\sigma_t}
e^{-\frac{t^2}{2S^2\sigma_t^2}}+
f_{tail}\frac{1}{\sqrt{2\pi}S_{tail}\sigma_t}
e^{-\frac{t^2}{2S_{tail}^2\sigma_t^2}},
\end{eqnarray*}
where $S$ and $S_{tail}$ are global scaling factors for the estimated error $\sigma_t$
for the main and tail Gaussian distributions and
$f_{tail}$ is a fraction of the tail part.
Fig.\ref{Fit-Results} shows the proper-time distributions and fit results
for $\dz\to\kpi$ and $\ds\to\phi\pip$.

%%\section{Systematic Uncertainties}
%%\noindent

\begin{table}[htbp]
\tcaption{Comparison of our results with PDG99\cite{PDG99} world average and previous measurements.}
\centerline{\footnotesize\smalllineskip
\begin{tabular}{l c c c c}\\
\hline
{} & $\tdz$~fs & $\tdp$~fs & $\tds$~fs & $\ycp$~\%\\
\hline
PDG99 & $415\pm4$                   & $1057\pm15$             & $495\pm13$               & --\\
E791  & $413\pm3\pm4$               & --                      & $(518\pm14\pm7)^\dagger$ & $0.8\pm2.9\pm1.0$\\
CLEO  & $408.5\pm4.1^{+3.5}_{-3.4}$ & $1034\pm22^{+10}_{-13}$ & $486\pm15\pm5$           & --\\
FOCUS & $409.2\pm1.3^\ddagger$      & --                      & $506\pm8^\ddagger$       & $3.42\pm1.39\pm0.74$\\
Belle & $\tdzresult$                & $\tdpresult$            & $\tdsresult$             & $\yresult$\\
\hline\\
\end{tabular}}
\vspace{-0.35cm}
{\scriptsize
\hspace{0.8cm}${}^\dagger$This result is included in the PDG99 world average.
${}^\ddagger$No systematic error is given.
}
\label{Comparison-Results}
\end{table}

%%\section{Conclusions}
%%\noindent
We measure the \Dz\ meson lifetime to be \Tdzresult\
using the decay mode $\dzkpi$.
The \Dp\ meson lifetime is measured to be ($1049 ^{+25}_{-24} {}^{+16}_{-19}$)~fs
for the $\dpkpipi$ decay sample and ($974^{+68}_{-62} {}^{+26}_{-18}$)~fs for
the $\dpphipi$ decay sample.
The \Ds\ meson lifetime is measured to be ($470 \pm 19 ^{+5}_{-7}$)~fs for the
$\dsphipi$ decay sample and ($505 ^{+34}_{-33} {}^{+8}_{-12}$)~fs for the
$\dskstk$ decay sample.
Table \ref{Comparison-Results} summarizes our combined measurement results with 
previous measurements and the world average.
The main sources of our systematic errors are uncertainties in the
resolution function, the proper-time dependence of the reconstruction
efficiency and a bias in the reconstruction of the decay vertex.
The ratios of the lifetimes of \Dp\ and \Ds\ with respect to \Dz\ are
measured to be \Rdpresult\ and \Rdsresult. 
The mixing parameter $\ycp$ is also measured through the lifetime
difference of \Dz\ mesons decaying into CP-mixed states and CP eigenstates.
We find \Yresult, corresponding to a 95\% confidence interval \Ylimit.

%%\nonumsection{Acknowledgements}
%%\noindent
We acknowledge 
the assistance of the staffs of KEK and of
all the participating institutions.
We acknowledge support from
the Ministry of Education, Science, Sports and Culture of Japan and
the Japan Society for the Promotion of Science.

\nonumsection{References}

\end{document}